\documentstyle[12pt]{article}
\begin{document}
\thispagestyle{empty}
\topmargin -1.0 cm
\headsep 1.5 cm
\textheight 22.0  cm
\normalbaselineskip = 24 true pt
\normalbaselines
\begin{flushleft}
{\large\sf SINP-TNP/96-01\\
TIFR/TH/96-01\\
January 1996}
\end{flushleft}

\begin{center}
{\Large\bf
Rho Mesons in Asymmetric Nuclear Matter:\\
A Renormalization Group Approach} 
\vskip 0.2 cm
{\sf Abhee K. Dutt-Mazumder \footnote{E-mail:
abhee@tnp.saha.ernet.in}, Triptesh De and Binayak Dutta-Roy}\\ 
Theory Group,\\
Saha Institute of Nuclear Physics,\\
1/AF  Bidhan Nagar, Calcutta - 700 064, India\\[1mm]
and\\[1mm]
{\sf Anirban Kundu \footnote{E-mail: akundu@theory.tifr.res.in}}\\
Theoretical Physics Group,\\
Tata Institute of Fundamental Research,\\
Homi Bhabha Road, Bombay - 400 005, India.
\end{center}
\begin{abstract}

The running of the $\rho\bar N N$ couplings in presence of
nuclear matter is studied, and the 
relevant $\beta$-functions are shown to be different for
$\rho^0$ and $\rho^+$, when the nuclear medium is asymmetric.
\end{abstract}

\newpage

The properties of dense nuclear matter and of particles propagating
in it have been seriously investigated in recent times \cite{je,sh}. 
Such studies are important from the viewpoint of nuclear astrophysics
as well as heavy ion collisions in the upcoming hadronic 
colliders \cite{ruivo,br}.
A proper development of the theory will undoubtedly help one to
analyze the signals obtained in these collisions and thus to determine
the equation of state of the system. Here we focus only on $\rho$
mesons inside a nuclear medium, as its significant role in heavy ion
collisions in connection with the dilepton spectroscopy has been 
pointed out earlier \cite{dls}. We show that the $\rho \bar N N$
coupling is not constant; rather, it is ``running" in nature and as
one increases the four-momentum $q$ of the $\rho$ meson, the value of
this coupling increases, analogous to what one sees in QED. The case of
asymmetric nuclear medium is more interesting: the $\rho^0\bar N N$
and the $\rho^+\bar N N$ couplings run in a different way, and if they
are same at some low value of $q^2$, they are bound to be different 
at a higher $q^2$ ($q$ always denotes the four-momentum of the $\rho$
meson). This effect vanishes for a symmetric medium, as expected.

We start with the Lagrangian \cite{sh} 
\begin{equation}
{\cal L}_{int} = g_v\bar N\gamma^{\mu}\tau_iN\rho^i_{\mu},
\end{equation}
where $\tau_i$s ($i=1$ to 3) are the usual $2\times 2$ Pauli matrices,
and $N$ is the two-component ($p$ and $n$) nucleon spinor. 
The Lagrangian in \cite{sh} contains a tensor interaction term too; 
however, we have not taken that into our subsequent discussions 
for simplicity. Effects of that term upon our results are discussed
later. For our 
later convenience, we write eq. (1) as
\begin{equation}
{\cal L}_{int} = g_v^0[\bar p\gamma^{\mu}p-\bar n\gamma^{\mu}n]\rho_{\mu}
^0+\sqrt{2}g_v^+[\bar p\gamma^{\mu}n \rho_{\mu}^+ + {\rm h.c.}]
\end{equation}
with $\rho^0\equiv \rho^3$ and $\rho^{\pm}\equiv (\rho^1\mp i\rho^2)/
\sqrt{2}$. 
The superscripts on the coupling constants are inserted merely to keep
track of the neutral and the charged meson interactions, which, though 
same in the Lagrangian, are anticipated to run differently in asymmetric
nuclear matter.

In nuclear medium, the nucleon propagator takes the form
\begin{equation}
G(k)=G_F(k)+G_D(k)
\end{equation}
where
\begin{equation}
G_F(k)={i(k\!\!\!/+M^{\ast})\over k^2-{M^{\ast}}^2+i\epsilon}
\end{equation}
and
\begin{equation}
G_D(k)=-(k\!\!\!/+M^{\ast})\Big[ {\pi\over E^{\ast}(k)} \delta
(k_0-E^{\ast}(k))\theta (k_F-|\vec k|)\Big]
\end{equation}
with $M^{\ast}$ and $E^{\ast}$ denoting, respectively, the effective mass
and total energy of the nucleon in the medium\cite{se}.
The second term, eq. (5),
comes from the Pauli blocking of nucleon states and the effective mass
of the nucleon is different from its mass in vacuum due to the presence 
of this Fermi sphere.

Before one embarks on the one-loop calculations, it is worthwhile to note
that the integrations having one or more $G_D$ are all finite; this is
because the momentum integration has a finite upper limit dictated by the
$\theta$-function.
Thus, only those terms which contain vacuum propagator for 
nucleon have divergences and are relevant for performing renormalization.
The rest of the paper will solely focus on these infinite terms.

One finds that the $\rho\bar N
N$ vertex renormalization $Z_1$ and the nucleon wavefunction 
renormalization $Z_2$ are equal, and using dimensional regularization
and setting the renormalization scale at $\mu _R$, one obtains
\begin{equation}
Z_1=Z_2=1-{1\over 8\pi^2}{1\over\epsilon}\Big({\mu\over\mu_R}\Big) ^
{\epsilon}[(g_v^0)^2+2(g_v^+)^2]
\end{equation}
where $\epsilon=4-d$, $d$ being the dimension in which the integrations
are performed. The $\rho^+\bar p n$ vertex 
contains wavefunction renormalizations for both proton and neutron; 
though the mass renormalizations for these two fermion fields are
different in an asymmetric nuclear medium, the wavefunction 
renormalizations are still same, which we denote as $Z_2$.
A pertinent question is the choice of $\mu_R$: for
$Z_2$, we take $\mu_R^2=k^2$, $k$ being the four-momentum of the
nucleon, and for $Z_1$, we again take $\mu_R=k$, this time $k$ being
the four-momentum of the nucleon coming in the vertex. It is clear that
this procedure is exactly analogous to QED; the only difference is that 
for QED, $Z_1$ is evaluated setting the four-momentum of the photon
to zero. The renormalized coupling contains $Z_1^{-1}Z_2$, and so these
two terms do not appear in the final formulae.

Next we proceed to find $Z_3$, the $\rho$-wavefunction renormalization.
The renormalized coupling $g_v^{ren}$ is related to the unrenormalized
coupling $g_v^{un}$ by
\begin{equation}
g_v^{ren}=Z_1^{-1}Z_2Z_3^{1/2}~g_v^{un}=Z_3^{1/2}~g_v^{un}.
\end{equation}
The two Feynman diagrams needed to compute $Z_3$ for $\rho^0$ and
$\rho^+$ are shown in figs. 1(a) and 1(b) respectively. For $\rho^0$, one
gets
\begin{equation}
i\Pi^{\mu\nu}(q^2)=i{(g_v^0)^2\over 3\pi^2}\Big({\mu\over\mu_R}\Big)^
{\epsilon}{1\over\epsilon}(q^{\mu}q^{\nu}-q^2g^{\mu\nu})
\end{equation}
and for $\rho^+$,
\begin{eqnarray}
i\Pi^{\mu\nu}(q^2)&=&i{(g_v^+)^2\over 3\pi^2}\Big({\mu\over\mu_R}\Big)^
{\epsilon}{1\over\epsilon}(q^{\mu}q^{\nu}-q^2g^{\mu\nu})\nonumber\\
&{ }&+i{(g_v^+)^2\over 2\pi^2}\Big({\mu\over\mu_R}\Big)^{\epsilon}
{1\over\epsilon}(M_p^{\ast}-M_n^{\ast})^2g^{\mu\nu}.
\end{eqnarray}

Note that $\Pi^{\mu\nu}$ in eq. (8) has the same Lorentz structure as
that of a bare propagator in Landau gauge. Borrowing the terminology
of QED, we can say that it is ``gauge invariant", though we stress 
that the symmetry discussed here is not a gauged one. This is because
$\rho^0$ is blind to the isospin projection of the nucleon in so far
as the $\Pi$-function is concerned, as the square of its coupling 
strength is same for both proton and nucleon, and both of them 
contribute equally. It does not matter whether the nuclear medium is 
asymmetric or not, since only the free part ($G_F$) comes into play.

This is, however, not true for $\rho^+$ (and $\rho^-$). As is evident
from eq. (9), an extra piece proportional to $g^{\mu\nu}$ gets added
to the usual $\Pi$-function (we again emphasize that only the divergent
parts are being focussed upon). This part comes from the breaking of
isospin symmetry in the nucleon sector ($M_p^{\ast}\not= M_n^{\ast}$),
and from pure dimensional arguments, is proportional to $(M_p^{\ast}
-M_n^{\ast})^2$. Obviously, for symmetric matter, this part will vanish,
and both $\rho^0$ and $\rho^{\pm}$ will have same $\Pi$-functions.

One can easily extract $Z_3$ from eq. (8); it is
\begin{equation}
Z_3^{\rho^0}=1-{(g_v^0)^2\over 3\pi^2}\Big({\mu\over\mu_R}\Big)^{\epsilon}
{1\over\epsilon}.
\end{equation}
Extraction of $Z_3$ from eq. (9) is a little tricky, because the 
coefficients of $g^{\mu\nu}$ and $-q^{\mu}q^{\nu}/q^2$ are not equal. 
Consider the analogous procedure for a gauge theory where the 
$\Pi$-function happens to be not gauge invariant. One then has, in the 
action term $F_{\mu\nu}F^{\mu\nu}$ for the gauge field, a part which
is proportional to $g^{\mu\nu}$ and only have the field renormalization
$Z_3$ in it (coming from $A^{\mu}\Box A^{\nu}$, where $A^{\mu}$ is the
gauge field and $\Box\equiv \partial^2/\partial t^2-\partial^2/\partial
{\vec x}^2$). The second part, which is proportional to $q^{\mu}q^{\nu}$
(stemming from $(\partial^{\mu}A_{\mu})^2$), contains both $Z_3$ as well
as the renormalization of the gauge-fixing term (the gauge-fixing term
needs renormalization {\em only} because the $\Pi$-function is not gauge 
invariant). Analogous to the above procedure, we collect the coefficient 
of $g^{\mu\nu}$ in eq. (9) and have
\begin{equation}
Z_3^{\rho^+}=1-{(g_v^+)^2\over 3\pi^2}\Big({\mu\over\mu_R}\Big)^
{\epsilon}{1\over\epsilon}\Big[1-{3(M_p^{\ast}-M_n^{\ast})^2\over 2q^2}
\Big].
\end{equation}
 
To remove the unphysical parameter $\mu$, we compare the values of $g_v$
at two different renormalization scales, {\em viz.}, $\mu_R$ and $\mu'_R$,
which gives
\begin{eqnarray}
g_v^0(\mu_R)&=&g_v^0(\mu'_R)-{(g_v^0)^3\over 6\pi^2}\Big({\mu'_R\over
\mu_R}\Big)^{\epsilon}{1\over\epsilon},\\
g_v^+(\mu_R)&=&g_v^+(\mu'_R)-{(g_v^+)^3\over 6\pi^2}\Big({\mu'_R\over
\mu_R}\Big)^{\epsilon}{1\over\epsilon}
\Big[1-{3(M_p^{\ast}-
M_n^{\ast})^2\over 2q^2}\Big].
\end{eqnarray}
\noindent From eqs. (12) and (13), 
we find the $\beta$-functions for $g_v^0$ 
and $g_v^+$ to be
\begin{eqnarray}
\beta(g_v^0)&\equiv &{\partial g_v^0\over\partial \ln\mu_R}
={(g_v^0)^3\over 6\pi^2},\\
\beta(g_v^+)&\equiv &{\partial g_v^+\over\partial \ln\mu_R}
={(g_v^+)^3\over 6\pi^2}
\Big[1-{3(M_p^{\ast}-
M_n^{\ast})^2\over 2q^2}\Big].
\end{eqnarray}
Writing $\alpha_v^0\equiv (g_v^0)^2/4\pi$ and 
$\alpha_v^+\equiv (g_v^+)^2/4\pi$, one obtains
\begin{eqnarray}
{1\over\alpha_v^0(Q^2)}&=&{1\over\alpha_v^0(\mu^2)}-{2\over 3\pi}
\ln {Q^2\over\mu^2},\\
{1\over\alpha_v^+(Q^2)}&=&{1\over\alpha_v^+(\mu^2)}-{2\over 3\pi}
\Big[1-{3(M_p^{\ast}-
M_n^{\ast})^2\over 2q^2}\Big]
\ln {Q^2\over\mu^2}.
\end{eqnarray}
Eqs. (16) and (17) give us the required running of $g_v$ with $q^2$. 
Of course, one needs to know the values of $\alpha_v^0$ and $\alpha
_v^+$ at a certain scale $\mu$, which need not be the same for 
$\rho^0$ and $\rho^+$, as we know \cite{we} that in an asymmetric
nuclear medium, the mass degeneracy of the $\rho$-isotriplet is
lifted.

Eq. (16) is quite analogous to QED in the sense that $\alpha_v^0$
increases with $q^2$, and the dependence, apart from numerical 
factors, has the same functional form. Taking as per the Bonn group
estimate \cite{bonn} $g_v^0=2.67$ at $\mu=0.77$ GeV, the mass
of $\rho^0$ in vacuum, we see from eq. (16) that this model
with $\rho^0$ and nucleon starts behaving like a free field theory 
from $q^2\approx 50 $ GeV, which is the well-known Landau pole.
On the other hand, the behaviour of $\alpha_v^+$ is interesting, more 
so because the coefficient of the logarithmic running term
is a function of $q^2$. The essential features are
similar for both the cases; an infrared (IR) fixed point exists at
$q^2=0$, and with increasing $q^2$, $\alpha_v$ hits the Landau pole.
However, the dominating terms for small $q^2$ are different and hence
the slope near the IR fixed point also differs.

We, however, note that in the stable collective 
excitation region, the Landau pole is never reached, and only a 
very limited portion of $\alpha_v(q^2)$ can be studied. Nevertheless,
we think that this study,  pursued in greater detail, can throw much light 
on the dynamics of low-energy hadronic systems.

One may ask what happens if we add a tensor interaction to the 
vector interaction taken in the present model. In that case, a ``gauge
invariant" piece, {\em i.e.}, a piece proportional to $q^{\mu}
q^{\nu}-q^2g^{\mu\nu}$, gets added to the $\Pi$-functions for $\rho^0$
and $\rho^+$. The magnitude of this term is, in general, quite small, 
and thus hardly any qualitative change in the $\beta$-functions takes
place.

\bigskip

To summarize, we have shown that the $\rho\bar N N$ couplings change 
with $q^2$ in a way similar to QED, and have an IR fixed point at the
origin. In asymmetric matter, the Lorentz structure of the 
$\Pi$-functions for $\rho^0$ and $\rho^+$ --- even the divergent 
parts only --- differ, due to the breaking of nuclear isospin
symmetry. This is a typical matter-induced feature and is absent 
for field theories in vacuum. The  model is shown to have a
Landau pole, and like QED, this pole is quite outside the physically 
interesting  region.

\bigskip

We thank V. Ravindran for useful discussions.

\newpage

\centerline{Figure captions}

{\bf 1(a)}.  Vacuum polarization diagram for $\rho^0$. Both $p\bar p$ and 
$n\bar n$ loops contribute. The loop momentum is $q$.

{\bf 1(b)}.  Vacuum polarization diagram for $\rho^+$. 

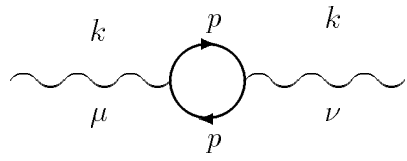
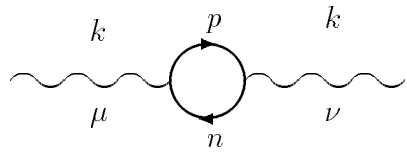
\end{document}